\documentclass[12pt,a4paper,epsf]{article}
\usepackage{epsf}
\usepackage{graphicx}

\newtoks\nslashfraction
\nslashfraction={.15}
\newcommand{\nslash}[1]{\setbox0\hbox{$ #1 $}
\setbox0\hbox to \the\nslashfraction\wd0{\hss \box0}/\box0 }

\newcommand{\lsim}{\buildrel < \over 
{_\sim}}

\newcommand{\be}{\begin{equation}}
\newcommand{\ee}{\end{equation}}
\newcommand{\bea}{\begin{eqnarray}}
\newcommand{\eea}{\end{eqnarray}}
\newcommand{\ba}{\begin{array}}
\newcommand{\ea}{\end{array}}

\begin{document}
\hfill{KRL MAP-298}

\vspace{0.5cm}

\begin{center}
{\Large {\bf The Vector Analyzing Power in 
Elastic Electron-Proton Scattering}}

\vspace{0.5cm}

\centerline{\sc L. Diaconescu$^{\;1}$ and M. J. 
Ramsey-Musolf$^{\;1,2}$}
\vspace{6pt}
\centerline{\it $^1$Kellogg Radiation 
Laboratory, California Institute of
                Technology, 
Pasadena, CA 91125, USA}
\vspace{6pt}
\centerline{\it $^2$Department of 
Physics,
                University of Connecticut, Storrs, CT 06269, 
USA}

\vspace{18pt}

\vspace{0.25in}
\vspace{0.25cm}
\vspace{0.25in}
\vspace{0.25cm}
\vspace{0.50in}

\begin{abstract}
We compute the vector analyzing power (VAP) for the elastic 
scattering of transversely polarized electrons from protons at low energies 
using an effective theory of electrons, protons, and photons. We study 
all contributions through second order in $E/M$, where $E$ and $M$ 
are the electron energy and nucleon mass, respectively.  The leading 
order VAP arises from the imaginary part of the interference of one- and 
two-photon exchange amplitudes. Sub-leading contributions are generated 
by the nucleon magnetic moment and charge radius as well as recoil 
corrections to the leading-order amplitude. Working to ${\cal O}(E/M)^2$, 
we obtain a prediction for $A_n$ that is free of unknown parameters and 
that agrees with the recent measurement of the VAP in backward angle 
$ep$ scattering. 

\end{abstract}

\vspace{0.30in}
pacs: 11.30.Er, 14.20.Dh, 25.30.Bf

\end{center}

\pagebreak

\section{Introduction}
\label{sec:intro}
\indent
\indent

The study of the vector analyzing power (VAP), $A_n$, in polarized 
electron-proton scattering has recently become a topic of considerable 
interest in nuclear physics. The VAP is a time-reversal (T) odd, parity (P) 
even  correlation between the electron spin and the independent momenta 
associated with the scattering process:
\be
\label{eq:ancorr}
A_n\sim \epsilon^{\mu\nu\alpha\beta}P_\mu S_\nu K_\alpha 
K^\prime_\beta\ \ \ ,
\ee
where $S$, $P$, and $K$ ($K^\prime$) denote the electron spin, initial 
proton momentum, and incident (scattered) electron momentum, 
respectively. A non-zero VAP cannot arise at leading order in quantum electrodynamics (QED), but could be generated by new T-odd, P-even interactions involving electrons and 
quarks. Searches for such interactions have been carried out in neutron and 
nuclear $\beta$-decay as well as nuclear $\gamma$-decays\cite{Lising:2000pa,Herczeg1995,Boehm1995}. Indirect constraints may also be obtained from limits on the permanent electric 
dipole moments of neutral atoms under various assumptions regarding the 
pattern of symmetry-breaking\cite{Kurylov:2000ub,Ramsey-Musolf:1999nk,Engel:1995vv,Conti:xn,Khriplovich:1990ef}. The sensitivity of direct searches for 
T-odd, P-even interactions is generally limited by the presence of QED 
\lq\lq final state interactions" (FSIs) that break the T-symmetry between initial 
and final states and give rise to non-vanishing T-odd, P-even 
observables. Uncertainties in theoretical calculations of these final state 
interactions would cloud the interpretation of a sufficiently precise T-odd, 
P-even measurement in terms of new interactions.  Observations of 
T-odd, P-even correlations in nuclear $\gamma$-decays are consistent with 
theoretical calculations of QED final state interactions\cite{Davis1980}, while T-odd, 
P-even searches in neutron $\beta$-decay have yet to reach the sensitivity needed 
to discern these effects.

Recently, the SAMPLE collaboration has reported a non-zero measurement 
of the VAP in polarized, elastic electron-proton scattering\cite{trans-ex}, making it 
the first non-zero result for any T-odd, P-even observable in any 
electron scattering process. The result has received widespread attention, 
as it differs substantially from the simplest theoretical estimate of 
QED final state contributions that neglects proton recoil and internal 
structure\cite{mott}. While one might speculate that this difference reflects the 
presence of new physics, a more likely explanation lies in elements of 
nucleon structure omitted from the simplest treatments of QED FSIs. 

If so, then the SAMPLE result, as well as other VAP measurements that 
have been completed or are under consideration, could have important 
implications for the interpretation of other precision observables 
involving hadrons that require computation of QED corrections to the leading 
order amplitude. Such observables include 
the ratio of proton electromagnetic form factors obtained via 
Rosenbluth separation in elastic $ep$ scattering\cite{Blunden:2003sp}, higher-order \lq\lq box graph" 
contributions to weak interaction observables\cite{McKeown:2002by}, or QED final state 
interactions in direct searches for T-odd, P-even effects. In each instance, 
a calculation of QED corrections requires a realistic and sufficiently 
precise treatment of hadronic intermediate states, particularly those 
arising in two-photon exchange amplitudes, ${\cal M}_{\gamma\gamma}$, or the analogous 
amplitudes involving the exchange of one heavy gauge boson and one photon. 
Since the leading QED contribution to $A_n$ arises from ${\rm Im}\ {\cal 
M}_{\gamma\gamma}$, experimental measurements of the VAP provide an 
important test of theoretical calculations of
${\cal M}_{\gamma\gamma}$ needed for the interpretation of other 
measurements.

At the same time, the VAP provides a new window on nucleon structure, 
as ${\cal M}_{\gamma\gamma}$ probes the doubly virtual Compton 
scattering (VVCS) scattering amplitude. 
In recent years, virtual Compton scattering (VCS) on the proton has 
become an important tool in probing the internal structure of the 
proton. VCS involves the 
coupling of one virtual and one real photon to a hadronic system. In 
the case of the proton,
the VCS cross section is sensitive to the generalized polarizabilities 
of the proton, and its measurement should provide insight in the proton structure 
\cite{guich}. In 
practice however, this
cross section includes Bethe-Heitler (BH) amplitudes associated with 
radiation of a real photon
from the electrons. Proper treatment of the cross section must 
therefore be taken in order to
obtain a correct interpretation of the measurement. In contrast, the 
process involving the 
coupling of two virtual photons to the hadronic system is immune to 
background BH amplitudes
and, thus, offers an alternative to VCS in probing the proton structure.

With the aforementioned motivation in mind, we study the VAP in the framework of an effective theory of low-energy $ep$ scattering. Since the SAMPLE measurement corresponds 
to kinematics close to the pion electroproduction threshold, we consider only the 
electron, photon, and nucleon as dynamical degrees of freedom. In this 
respect, our analysis corresponds to the use of heavy baryon chiral 
perturbation theory with the pions integrated out. To make the treatment 
systematic, we expand $A_n$ in powers of $p/M$, where $p$ is either the 
incident electron energy ($E$) or mass ($m$) and $M$ is the nucleon mass.  Working 
to second order in $p/M$, we obtain all contributions to $A_n$ that 
arise uniquely from one-loop, two-photon exchange amplitudes and obtain 
a prediction that is free from any unknown parameters.   We also write 
down the leading, non-renormalizable T-odd, P-even $eepp$ operators whose 
intereference with ${\cal M}_\gamma$ can generate a non-zero VAP and show that 
they contribute at ${\cal O}(p/M)^4$. 

We find that inclusion of all one-loop effects through ${\cal 
O}(p/M)^2$ in ${\cal M}_{\gamma\gamma}$ as well as all terms in ${\cal M}_\gamma$ through this order is sufficient to resolve the disagreement between the SAMPLE result and the 
simplest potential scattering predictions. This resolution follows from 
several effects that occur beyond leading order in $p/M$: recoil corrections 
to the pure charge scattering result obtained in Ref. \cite{mott},  the 
nucleon isovector magnetic moment, and the proton charge radius.
In the absence of dynamical pions, contributions from the nucleon polarizability
arise at higher order than we consider here and appear unnecessary to account for the experimental result. Given that 
the incident electron energy $E$ is of the same order as $m_\pi$,
we have no {\em a priori} reason to expect agreement of our computation 
with experiment. What it suggests, however, is that for this kinematic 
regime, pions play a less important role in the VVCS amplitude than one might naively expect.   
Future, low-energy $A_n$ measurements, taken over a broader range in 
$q^2$ and scattering angle than relevant to the SAMPLE measurement, would 
provide additional, useful tests of this conclusion.

We also consider $A_n$ at forward scattering angles and energies 
somewhat higher than those of the SAMPLE experiment, since preliminary 
results for this kinematic domain have been reported by the A4 Collaboration 
at the MAMI facility in Mainz\cite{Maas2003}. Although we would not expect our 
framework to be reliable in this kinematic regime, where the electron energy 
$E$ is much closer to $M$, it is nonetheless instructive to compare 
with the Mainz preliminary results as a way of pointing to the physics 
that may be operative in this domain. Indeed, we find substantial 
disagreement with the preliminary Mainz data. The culprit could be that going to the Mainz 
kinematics  exceeds the limit of validity of our effective theory, that 
we must  include additional dynamical degrees of freedom such as the 
$\pi$ or $\Delta (1230)$ resonance, or both. Future studies using 
alternative methods such as dispersion relations may be needed to 
explore this kinematic domain.

Finally, we also consider $A_n$ for polarized M\o ller scattering. The VAP for this process has been measured by the E158 Collaboration at SLAC \cite{E158}, and theoretical computations given  in Refs. \cite{Barut:1960,DeRaad:1974,Dixon2004}. Our computation agrees with these earlier $A_n(ee)$ calculations, providing a useful cross-check on our study of the VAP for $ep$ scattering. 

Our discussion of these points is organized in the remainder of the 
paper as follows. In Section 2, we discuss general features of $A_n$ and 
our approach to the computation. Section 3 provides details of the 
calculation. In Section 4, we give numerical results and discuss their 
significance, while Section 5 gives our conclusions. Technical details are 
provided in the Appendices.

\section{General Considerations}
\label{sec:general}
\indent
\indent
We are interested in computing the VAP in elastic $ep$ scattering:
\bea
\label{eq:an}
A_n &=& 
{d\sigma_{\uparrow}-d\sigma_{\downarrow}\over d\sigma_{\uparrow}+d\sigma_{\downarrow}}
= {2{\rm Im} \ {\cal M}_{\gamma\gamma}^\ast{\cal M}_{\gamma}\over 
\vert{\cal M}_{\gamma}\vert^2}\ \ \ ,
\eea
where $d\sigma_{\uparrow(\downarrow)}$ is the differential cross section for 
scattering of electrons with incident spin parallel (anti-parallel) to ${\vec 
K}\times{\vec K^\prime}$. 
In a phase convention where the single $\gamma$-exchange amplitude 
${\cal M}_{\gamma}$
is purely real, $A_n$ requires a non-vanishing imaginary part of ${\cal 
M}_{\gamma\gamma}$\footnote{By ${\rm Im}{\cal M}_{\gamma\gamma}$, we mean the coefficients of the various products of fermion bilinears, ${\bar e}\Gamma e {\bar N}\Gamma^\prime N$, {\em etc.} that appear in the amplitude.} .
To compute the latter, one must consider both the box and crossed-box 
diagrams of Fig. 1. Simple power-counting arguments indicate that the 
contribution to ${\cal M}_{\gamma\gamma}$ arising from the leading-order 
$\gamma p$ couplings is ultraviolet finite but infrared divergent. 
Thus, in general, one must also compute the contributions to $A_n$ 
arising from the bremsstrahlung diagrams of Fig. 2. As we show by explicit 
calculation in Appendix A, however, the bremsstrahlung contribution to 
$A_n$ vanishes identically, while ${\rm Im} {\cal M}_{\gamma\gamma}$ is infrared 
finite. The resulting, leading-order contribution to $A_n$ is ${\cal O}(p/M)^0$.
\begin{figure}
    \begin{center}
    \includegraphics[angle=90,width=9cm]{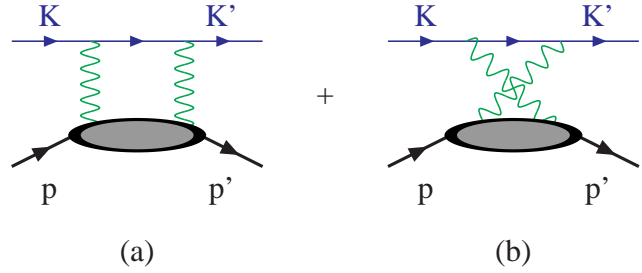}
    \end{center}
    \caption{(Color online) Two photon exchange diagrams. The wavy lines indicate virtual photons, while $k$ ($k^\prime$) and $p$ ($p^\prime$) denote the initial (final) electron and proton momenta, respectively.}   
\end{figure}

Additional contributions to ${\cal M}_{\gamma\gamma}$ arise from 
higher-order operators that couple one or more virtual photons to the proton 
and electron. We neglect the latter since they are suppressed by 
additional powers of the fine structure constant\footnote{For high energy scattering, these higher-order QED contributions may receive logarithmic enhancements\cite{Dixon2004}.}. In contrast, the $\gamma p$ operators are 
induced by strong interactions and have couplings of order $e$. In order 
to treat their contributions systematically, we adopt an effective 
theory framework since we cannot compute the operator coefficients from 
first principles in Quantum Chromodynamics. The natural framework for 
doing so is heavy baryon chiral perturbation theory (HB$\chi$PT), which 
provides a systematic expansion in powers of $p/\Lambda_\chi$ and $p/M$, 
where $\Lambda_\chi=4\pi F_\pi$ is the scale of chiral 
symmetry-breaking and $p$ is an external momentum or mass with magnitude much less than 
$M$ and $\Lambda_\chi$. In the present case, where we integrate out 
the pions, we take $p=$ $E$ or $m$ and use $M$ as the heavy scale.  For the kinematics of the SAMPLE experiment, $E>> m$. Since there are no hard collinear infrared 
singularities in ${\rm Im}\ {\cal M}_{\gamma\gamma}$,  we may drop all power 
corrections involving the electron mass and obtain our result as an 
expansion in $E/M$. 

\begin{figure}
    \begin{center}
    \includegraphics[angle=90,width=11cm]{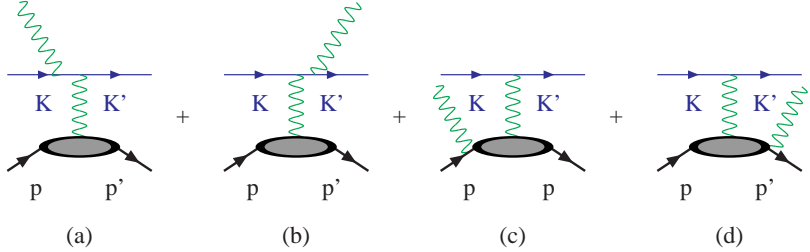}
    \end{center}
    \caption{(Color online) Bremsstrahlung contributions. Labels are the same as in Fig. 1.}   
\end{figure}
The leading terms in heavy baryon Lagrangian for nucleons and photons 
relevant to our computation are
\bea
\nonumber
{\cal L}_{N\gamma} &=& {\bar B}_v i v\cdot D B_v  +\frac{1}{2 M}{\bar B}_v \left[(v\cdot D)^2-D^2\right]B_v\\
\label{eq:lngamma}
&&+\frac{e\mu}{2M}\epsilon_{\mu\nu\alpha\beta}F^{\mu\nu} v^\alpha {\bar B}_v S^\beta B_v
-\frac{eC_r}{M^2}{\bar B}_v v_\mu B_v \partial_\lambda F^{\mu\lambda}+\cdots
\eea
where $B_v$ is the field for a heavy proton of velocity $v_\mu$, where  
$D_\mu =\partial_\mu-ieA_\mu$, and where we have shown explicitly all 
$\gamma p$ interactions through ${\cal O}(p^3)$. The latter arise from the 
subleading kinetic term in Eq. (\ref{eq:lngamma}) as well as from the operators containing the
field strength, $F^{\mu\nu}$. The coefficient $\mu = 2.793$ is the proton magnetic moment, while $C_r$ determines the proton Sachs, or electric, radius:
\be
C_r=\frac{M^2}{6}\langle r^2\rangle_E = M^2\frac{d G_E^p(t)}{dt}\vert_{t=0}\ \ \ ,
\ee
where $t=q^2$.  The experimental value for $\langle r^2\rangle_E=0.743$ fm$^2$ \cite{Simon:hu,Bhaduri} implies
$C_r=2.81$.
When included in the 
loop diagrams of Fig. 1, these interactions generate contributions to the 
$ep$ amplitude ${\cal M}_{\gamma}$ and ${\cal M}_{\gamma\gamma}$ 
through order 
$(p/M)^2$ relative to the leading term. To this order, operators 
associated with the nucleon polarizability (see Fig. 3e) do not contribute, as they occur 
at ${\cal O}(p^4)$ in ${\cal L}_{N\gamma}$ when the pion is treated as 
heavy. 

Higher-order contributions to $A_n$ can also arise from effective T-odd, P-even 
$eeNN$ interactions. The origin of such operators could be either physics 
that we have integrated out, such as contributions to ${\cal M}_{\gamma\gamma}$ from $\pi N$ or $\Delta$ intermediate states , or explicit T-odd, 
P-even interactions arising from new physics. As shown in Appendix B, 
there exist no Hermitian, four-fermion operators at dimension six that 
contribute to $A_n$. The lowest dimension T-odd, P-even four fermion operators have dimension seven and would nominally contribute to $A_n$ at ${\cal O}(p/M)^3$. We show, however, that contributions from these operators vanish to this order and first arise at ${\cal O}(p/M)^4$. Since we truncate our analysis at two orders lower, we may neglect these operators and obtain a parameter-free prediction for the VAP. Nevertheless, we discuss these operators briefly in Section 4 when considering the possible size of  neglected, higher-order contributions\footnote{For an earlier, phenomenological calculation that includes some of these higher order contributions, see Ref. 
\cite{DeRujula:1972te}.}. 

\begin{figure}
    \begin{center}
    \includegraphics[angle=90,width=11cm]{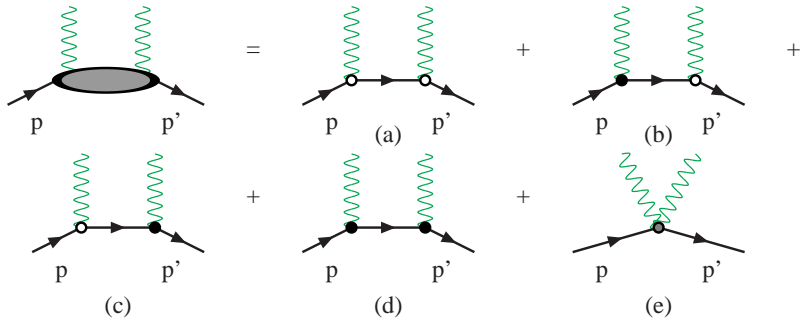}
    \end{center}
    \caption{(Color online) Contributions to the VVCS amplitude appearing in Fig. 1. Open circles indicate the leading order $\gamma N$ couplings, while dark circles indicate higher-order couplings, such as the magnetic moment and charge radius. Shaded circle denotes that nucleon polarizability operator.}   
\end{figure}

As we show in detail in Section 3, the leading one-loop contributions 
to $A_n$ -- generated by two ${\cal O}(p)$ $\gamma p$ insertions in the VVCS amplitude (Fig. 3a) -- are finite, non-analytic in $p$, and occur ${\cal O}(p/M)^0$, 
whereas those generated by the dimension seven T-odd, P-even operators  
arise at ${\cal O}(p/M)^4$. Thus, the leading contributions are uniquely 
determined from the one-loop calculation. Similarly, contributions to 
${\cal M}_{\gamma\gamma}$ involving one ${\cal O}(p)$ and one ${\cal 
O}(p^2)$ $\gamma p$ interaction (Fig. 3b,c) contribute to $A_n$ at ${\cal O}(p/M)$, are 
also finite and non-analytic in $p$, and are unique to the loop 
calculation. The ${\cal O}(p/M)^2$ loop contributions arise either from two 
${\cal O}(p^2)$ $\gamma p$ operators ({\em e.g.}, two insertions of the 
nucleon magnetic moment operator, Fig. 3d) or one ${\cal O}(p)$ and one ${\cal 
O}(p^3)$ term ({\em viz}, the proton charge radius). We find, however, that 
the ${\cal O}(p/M)^2$ components of ${\cal M}_{\gamma\gamma}$  arise only 
from the $\gamma p$ magnetic moment interaction as well as from recoil 
order terms in ${\cal L}_{N\gamma}$. Contributions to ${\cal 
M}_{\gamma\gamma}$ from the proton charge radius vanish, though it does 
contribute to $A_n$ 
as a higher-order term in ${\cal M}_\gamma$.

\section{Two-photon exchange}
\label{sec:twophoton}

The evaluation of four point functions for general kinematics does not 
readily lend itself to evaluation using standard Feynman 
parameterization in the loop integrals. Alternate methods for evaluating these 
integrals that do not rely explicitly on Feynman parameters have been worked 
out in Refs. \cite{thft,pave} and have become standard. In the 
present case, where we are interested in backward angle scattering at 
nonzero $q^2$, we would ideally like to use this formalism. However, the 
form of the heavy baryon propagator does not permit one to adopt the t'Hooft-Passarino-Veltmann formulation directly. 

We circumvent these difficulties by carrying out the computation with 
relativistic baryon propagators and expanding our result in powers of 
$p/M$. Doing so allows us to evaluate the loop integrals using the 
standard formulation of Refs. \cite{thft,pave}. It has been shown in other contexts\cite{Zhu:2000zf} that doing so allows one to recover the heavy baryon result so long as 
the external momenta are sufficiently small. Moreover, our loop
results are entirely non-analytic in $p$ and, thus, must match the 
corresponding non-analytic results obtained with heavy baryon propagators. To 
the order of our analysis, there exist no four fermion operators that could account 
for differences between relativistic and non-relativistic treatments of 
$A_n$. 

The one-loop ${\cal M}_{\gamma\gamma}$ is nominally infrared singular 
and must, therefore, be regulated with an IR regulator such as a photon 
mass. On general grounds, the regulator dependence should be cancelled by a 
corresponding dependence of the bremsstrahlung contribution to the 
spin-dependent cross section. As is well known, such a cancellation occurs for 
unpolarized scattering cross section. In Appendix B, we work out the 
corresponding bremsstrahlung contribution to $A_n$ and show that it 
vanishes identically. Consequently, ${\rm Im}{\cal M}_{\gamma\gamma}$ must 
be IR regulator-independent. 

In general,  the amplitude  ${\cal M}_{\gamma\gamma}$ depends on each 
of the eleven integrals obtained in Ref. \cite{pave}. The 
imaginary part, however, depends on only four:
\bea
D_0 &=& \frac{2\pi}{-t} 
\ln(\frac{-t}{\lambda^2})\frac{1}{\sqrt{\Lambda}}\Theta(s-(m+M)^2)  \nonumber \\
C_0(1,2,3) &=& \frac{\pi}{\sqrt{\Lambda}} 
\ln(\frac{\Lambda}{s\lambda^2})\Theta(s-(m+M)^2)  \nonumber \\
C_0(1,3,4) &=& C_0(1,2,3) = C_0 \nonumber \\
B_0(1,3) &=& \pi\frac{\sqrt{\Lambda}}{s}\Theta(s-(m+M)^2)
\eea
where the three labels associated with the $B_0$ and $C_0$ functions indicate which 
propagators are used for the two-point and
three-point integral as discussed in Appendix C, $\lambda$ is the photon mass, and
\bea
 \Lambda = s^2-2s(M^2+m^2)+(M^2-m^2)^2  
\eea
These integrals have been previousely computed in Refs. \cite{pave,vann} (In \cite{vann}
they are obtained by the use of dispersion techniques). The $D_0$ and $C_0$ 
loop integrals diverge as $\lambda\to 0$, but  the combination
\begin{eqnarray}
\label{eq:finite}
2C_0+D_0t = 
\frac{2\pi}{\sqrt{\Lambda}}\ln(\frac{\Lambda}{-st})\Theta(s-(m+M)^2) 
\end{eqnarray}
is finite in this limit and is the only combination of $D_0$ and $C_0$
integrals that is so. As such, 
the two-photon contribution to $A_n$ must only contain terms proportional to this 
combination or to the $B_0$ integral.

In evaluating the loop contributions to $A_n$, it is most efficient to 
identify the terms in ${\cal M}_{\gamma\gamma}$ that generate the 
correlation of Eq. ({\ref{eq:ancorr}) by carrying out the Dirac algebra in 
the interference term ${\rm Im}{\cal M}_{\gamma\gamma}\ {\cal M}_{\gamma}^\ast$ before 
evaluating the momentum integrals\footnote{This procedure introduces no 
ambiguities because ${\rm Im}\ {\cal M}_{\gamma\gamma}$ is finite to the 
order of our analysis.}. After carrying out the momentum integration, 
the contribution from the box diagram of Fig. 1a is
\bea
\label{eq:full}
2 {\rm Im}{\cal M_{\gamma\gamma}^{\rm box}}{\cal M}_{\gamma}^{\ast}
&=&-\frac{(4\pi\alpha)^2}{4\pi^4t}
\frac{16m\pi^2 
(4\pi\alpha)}{(\Lambda+st)}\epsilon^{\mu\nu\alpha\beta} P_\mu S_\nu K_\alpha K^\prime_\beta\nonumber \\
&\bigg\{&\bigg[4(M^2-m^2-3s)M^2R + \kappa [(6R+2)\Lambda-((m^2-M^2-s)R+2s)t] \nonumber \\
&+&\kappa^2R\frac{1}{8M^2(\Lambda+st)}[2(3m^3+16M^2)\Lambda^2 
\nonumber \\
&+&\Lambda(11m^4-2(13M^2+8s)m^2+15M^4+11s^2+14M^2s)t \nonumber \\
&+&4s(2m^4-(5M^2+4s)M^2+3M^4+2s^2-3M^2s)t^2]\bigg](2C_0+D_0t) \nonumber \\
&-&4\frac{\Lambda+ts}{\Lambda}(\kappa^2+4\kappa+2)B_0\bigg\}
\eea
$s$, $t$,  and $u$ are the Mandelstaam variables, $\kappa=\mu-1$ is the nucleon 
anomalous magnetic moment and 
\be
R-1= t \left[ \frac{\kappa}{4M^2}-\frac{C_r}{M^2}\right]\ \ \ .
\ee 

To obtain the result consistent with our power counting, we expand Eq. 
(\ref{eq:full}) in powers of $p/M$ up to second order relative to the 
leading term:
\bea
\label{eq:trunc}
{\rm Im}{\cal M_{\gamma\gamma}^{\rm box}}{\cal 
M}_{\gamma}^{\ast} &=&-\frac{(4\pi\alpha)^2}{t}\frac{32\pi^2\alpha m M}{\sqrt{E^2-m^2}[(E^2-m^2+t/4)+
\frac{E t}{2M}+\frac{m^2t}{4M^2}]}
\epsilon^{\mu\nu\alpha\beta}P_\mu S_\nu K_\alpha K^{\prime}_{\beta}\nonumber \\
&\times&\Bigg\{\Bigg[\ln\left[\frac{4(E^2-m^2)}{-t}\right]-2E/M+(2E^2-m^2)/M^2\Bigg]\nonumber \\
&\bigg[&R 
+\frac{3E}{M} +\frac{2m^2}{M^2}
+ \frac{\kappa^2}{M^2}\frac{32(E^2-m^2)^2+t^2/2+10(E^2-m^2)t}{4(E^2-m^2)+t} \nonumber \\
&+& \frac{4\kappa}{M^2}(m^2-E^2)\bigg] -\frac{\kappa^2+4\kappa+2}{M^2}\bigg[(E^2-m^2)+\frac{t}{4}\bigg]\Bigg\} \nonumber \\
&\Theta&(s-(m+M)^2)\ \ \ 
\eea
where the $\Theta$-function arises from the integrals $2C_0+2 D_0 t$ and $B_0$. 
Note that we have retained the $m$-dependence purely for illustrative 
purposes, as $m << E$ for the experiments of interest here. The 
corresponding contribution from the crossed-box diagram can be obtained by 
crossing symmetry with the replacement $s\to u$. In this case, the 
$\Theta$-function vanishes, so only ${\rm Im}{\cal M}_{\gamma\gamma}^{\rm box}{\cal M}_\gamma^{\ast}$ 
contributes. 

In the expression (\ref{eq:trunc}), the terms that go as powers of 
$E/M$ or $m/M$ but do not contain factors of $\kappa$ or $C_r$ arise purely from 
recoil effects. The proton charge radius contributes solely via ${\cal M}_\gamma$. 
Although it also contributes to the absorptive part of ${\cal M}_{\gamma\gamma}$, the resulting terms do not contribute to the spin-dependent correlation of Eq. (\ref{eq:ancorr}). 
Including the magnetic moment, charge radius, and 
recoil-order terms in ${\cal M}_{\gamma}$ along with the loop 
contributions in Eq. (\ref{eq:trunc}) leads to the following expression for the 
VAP:
\bea
\label{eq:vapfinal}
A_n&=&-\frac{2\alpha tm}{\sqrt{E^2-m^2}[(E^2-m^2+t/4)+
\frac{E t}{2M}+\frac{m^2t}{4M^2}]}
\vec{S}\cdot \vec{K} \times  \vec{K^{\prime}} \nonumber \\
&\times&\Bigg\{ \Bigg[ \ln\left[\frac{4(E^2-m^2)}{-t}\right]-2E/M+(2E^2-m^2)/M^2\Bigg]\nonumber \\
&\bigg[&R 
+\frac{3E}{M} +\frac{2m^2}{M^2}
+ \frac{\kappa^2}{M^2}\frac{32(E^2-m^2)^2+t^2/2+10(E^2-m^2)t}{4(E^2-m^2)+t}
+\frac{4\kappa}{M^2}(m^2-E^2)\bigg]\nonumber \\
&-&\frac{\kappa^2+4\kappa+2}{M^2}\bigg[(E^2-m^2)+\frac{t}{4}\bigg]\Bigg\} \nonumber \\
&\times&\bigg{[}(8E^2+2t)R^2 + \frac{4Et}{M} +t\frac{t+2m^2+2\kappa(t+2m^2)+\kappa^2
[t+4(m^2-E^2)]/2}{M^2}\bigg{]}^{-1}
\eea
Dropping all terms that go as powers of $E/M$, $m/M$, or $t/M^2$
yields the result obtained in Ref. 
\cite{mott} that was obtained for scattering from an  infinitely heavy, 
point-like proton.

\section{Results and Discussion}
\label{sec:results}

The expression for $A_n$ given in Eq. (\ref{eq:vapfinal}) provides a 
parameter-free prediction for low-energy electron scattering. In Fig. 4 and Fig. 5, 
we plot $A_n$ as a function of energy for fixed laboratory frame scattering angles 
$\theta=146.1^\circ$ (Fig. 4) and $\theta =30^\circ$ (Fig. 5), while 
in Fig. 6 we show the VAP for fixed energy $E=192$ MeV while 
varying $\theta$. In call cases, the leading order calculation is shown for comparison. In Fig. 6, the relative importance of the recoil, magnetic moment, and charge radius contributions are also indicated.

\begin{figure}
    \epsfxsize=4.0in
    \begin{center}
    \epsffile{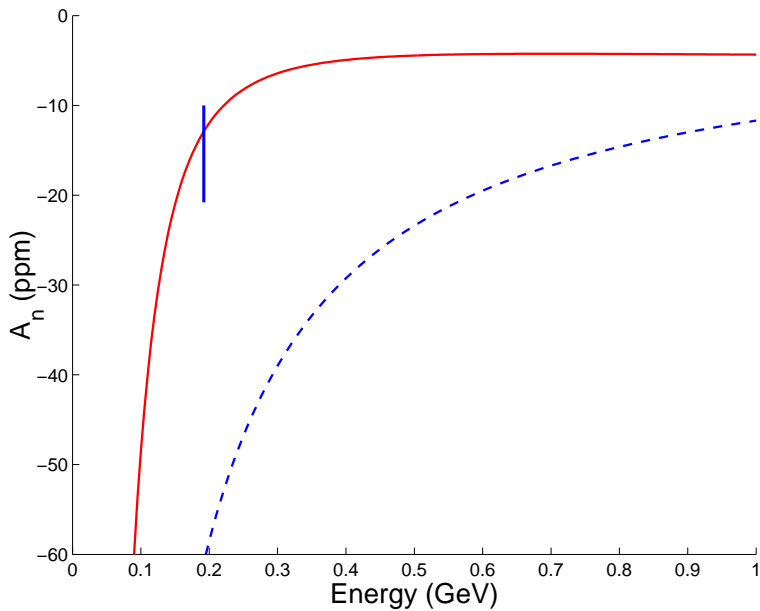}
    \end{center}
    \caption{(Color online) VAP vs energy for fixed scattering angle, $\theta=146.1^\circ$. The dashed blue line is the leading order result, and the solid red line shows the full calculation. The SAMPLE result\cite{trans-ex} is also shown at $E=192$ MeV.}
\end{figure}

\begin{figure}
    \epsfxsize=4.0in
    \begin{center}
    \epsffile{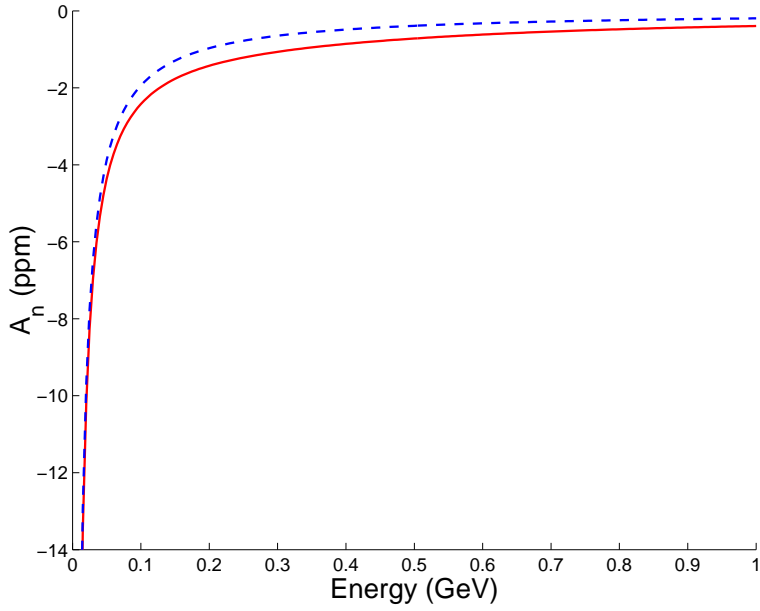}
    \end{center}
    \caption{(Color online) VAP vs energy for fixed scattering angle,   $\theta =30^\circ$. The dashed blue line is the leading order result, and the solid red line shows the full calculation.}
\end{figure}

The result obtained in the SAMPLE measurement is also shown. While the 
leading order calculation over estimates the magnitude of $A_n$ by a 
factor of roughly four, inclusion of the higher-order terms considered here 
produces agreement with the experimental value. Interestingly, there appears 
to be scant evidence that dynamical pions or the $\Delta$ play a 
significant role in $A_n$ for this kinematic region ($E=192$ MeV), despite one's 
expectation that they might.

\begin{figure}
    \epsfxsize=4.0in
    \begin{center}
    \epsffile{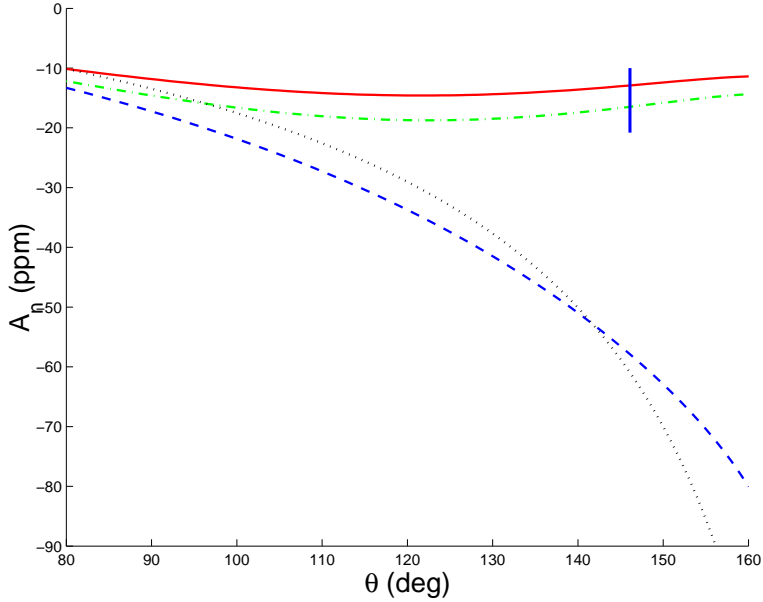}
    \end{center}
    \caption{(Color online) VAP vs scattering angle for the SAMPLE kinematics ($E=192$ MeV). 
The dotted black line gives the leading order result, the blue dashed line adds the recoil
corrections, the green dash-dotted line adds the magnetic corrections,and the solid red line shows the full calculation through ${\cal O}(p/M)^2$.}
\end{figure}

At higher energies, our result for $A_n$ cannot be considered reliable, 
since the convergence of the effective theory expansion breaks down for 
$E\sim M$. The A4 collaboration at Mainz has measured $A_n$ at 
$E=570.3$ MeV and $E=854.3$ MeV and $25^\circ\leq \theta \leq 35^\circ$.  Preliminary results for the higher energy VAP have been reported in Ref. \cite{Maas2003}. A comparison with our computation indicates that the preliminary experimental values for forward angle scattering and higher energies are 
substantially larger in magnitude than we are able to obtain via the 
low-energy expansion to ${\cal O}(E/M)^2$. Presumably, a resummation of higher-order 
contributions in $E/M$ using non-perturbative techniques, such as dispersion 
relations, would be required to compute reliably  $A_n$ in this 
domain\cite{Blunden:2003sp,Afanasev:2002gr,Chen:2004tw,Pasquini:bw,Bernabeu}. We would also expect that inclusion of nucleon resonances\footnote{For recent studies that pertain to such contributions, see Refs. \cite{Pasquini:2004pv,Afanasev:2004hp,Afanasev:2004pu}.} and 
pions as explicit degrees of freedom would be needed to account for the 
experimental results.

\begin{figure}
    \epsfxsize=4.0in
    \begin{center}
    \epsffile{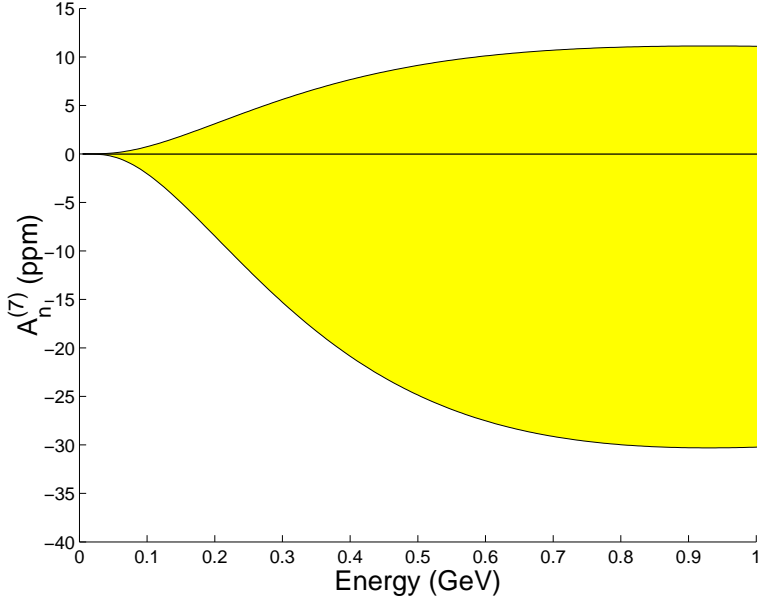}
    \end{center}
    \caption {(Color online) Possible contribution from the dimension seven, T-odd, P-even operator ${\cal O}_{eN}^{7a}$ to the backward angle VAP ($\theta=146.1^\circ$).}
\end{figure}
One indication of the possible strength of these higher-order contributions may be given by considering the T-odd, P-even dimension seven operators. As shown in Appendix B, there exist two $d=7$ operators that could, in principle, contribute. From an explicit calculation, we find that only one of the two -- ${\cal O}_{eN}^{7a}$ -- leads to a non-vanishing $A_n$.  Here, it is useful to consider the form of this operator for relativistic proton fields, $N$:
\be
\label{eq:dimsevenrel}
{\cal O}_{eN}^{7a}= \frac{\alpha^2 C_{7a}}{M^3} {\bar e}\sigma^{\mu\nu}\gamma_5 ({\overrightarrow D}+{\overleftarrow D} )_\nu e\ {\bar N}\gamma_5 \gamma_\mu N\ \ \ .
\ee
Re-writing this operator in terms of the heavy fields $B_v$ leads to
\be
{\tilde {\cal O}}_{eN}^{7a}= -2\frac{\alpha^2 C_{7a}}{M^2} {\bar e}\sigma^{\mu\nu}\gamma_5 ({\overleftarrow D}+{\overrightarrow D} )_\nu e\ {\bar B}_v S^v_{\mu} B_v\ \ \ ,
\ee
where $S^v_{\mu} $ is the nucleon spin. The contribution from ${\tilde {\cal O}}_{eN}^{7a}$ to the interference amplitude ${\rm Im}\ {\tilde {\cal M}}_{eN}^{7a}\ {\cal M}_\gamma^{\ast}$ goes as 
$\epsilon^{\mu\nu\alpha\beta} S_\mu v_\nu v_\alpha K^{\prime}_\beta$ and, thus, vanishes. On the other hand, using the relativistic form of the operator, ${\cal O}_{eN}^{7a}$, leads to the correlation $\epsilon^{\mu\nu\alpha\beta} S_\mu P_\nu P^{\prime}_\alpha K^{\prime}_\beta$ that is non-vanishing for 
$P\not= P^\prime$. The resulting contribution to the VAP is
\be
\label{eq:higher}
A_n^{(7)} = \frac{\alpha C_{7a}}{4\pi}{t^2 |\vec{K}| |\vec{K^\prime}| \sin\theta\over M^2[8M^2E^2+2(2E+M)tM+t^2]} \ ,
\ee
a result that is ${\cal O}(p/M)^4$. In short, the only heavy baryon operators that can contribute involve either fields with two different velocities ({\em viz}, $B_v$ and $B_{v'}$) whose contribution requires non-zero proton recoil, or dimension eight operators involving the $B_v$ fields only and carrying an additional $p/M$ recoil suppression.  

The SAMPLE result for $A_n$ allows for a non-vanishing, but small coefficient for the leading, higher-order T-odd, P-even operator. Using the relativistic operator 
${\cal O}_{eN}^{7a}$ for illustration and including the loop contributions through ${\cal O}(p/M)^2$ leads to $C_{7a}=3.07\pm 6.64$. Naive dimensional analysis would have suggested 
a magnitude for $C_{7a}$ or order unity, so the SAMPLE results do not appear to imply the presence of any un-natural hadronic scale physics. We may now use this range for $C_{7a}$ to estimate 
the possible size of higher-order effects at other kinematics. 
The resulting band is shown in Fig. 7 for backward angles ($\theta=146.1^\circ$) and in Fig. 8 for forward angles ($\theta=30^\circ$). For the Mainz measurement at $E=570$ MeV and $\theta = 30^\circ$, we find $-2.0 \leq A_n^{(7)} \leq 0.7$ ppm, while
$A_n^{\rm loop} = - 0.64$ ppm. Thus, one might expect the impact of the physics we have integrated out to grow in importance relative to the loop effects considered here as the energy of the beam is increased, and it appears reasonable to expect a magnitude of a few ppm at the Mainz kinematics. We caution, however, that the  precise value obtained in our calculation is unlikely to be correct in this energy regime, where the convergence of the $E/M$ expansion is slow at best. 

\begin{figure}
    \epsfxsize=4.0in
    \begin{center}
    \epsffile{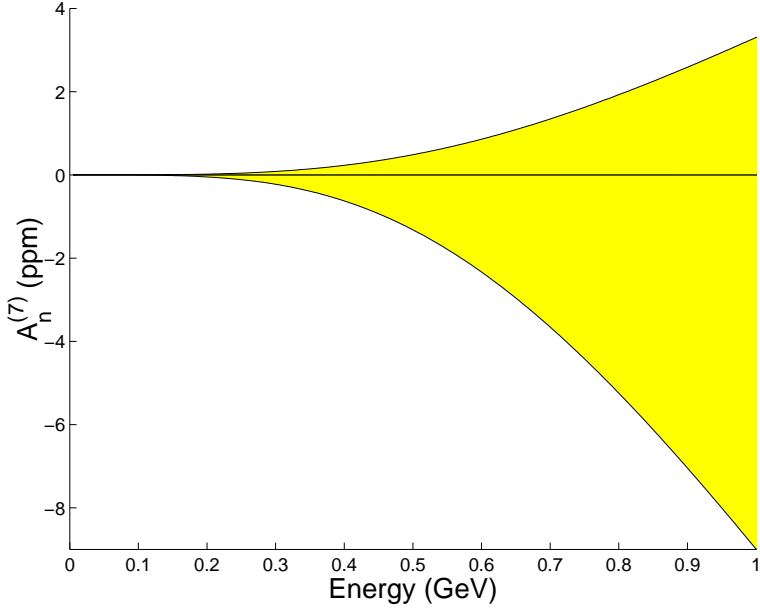}
    \end{center}
    \caption{(Color online) Possible contribution from ${\cal O}_{eN}^{7a}$ to the VAP at $\theta=30^\circ$, given constraints on the operator coefficient $C_{7a}$ implied by the SAMPLE result.}
\end{figure}

As a final comparison, we also consider $A_n$ in fixed target, 
polarized M\o ller scattering. The VAP for this process has been measured at SLAC 
by the E158 collaboration\cite{E158}, and one expects results to be forthcoming in the near future.
Calculations of this quantity have been performed by several authors\cite{Barut:1960,DeRaad:1974,Dixon2004}. As a cross-check on our VAP for $ep$ scattering, we carry out the analogous calculation here. It 
can be performed completely 
relativistically without performing an expansion in electron energy. However, since we are now 
dealing with identical particles in the final state we need to compute the interference between 
tree diagams in Figure 9b and the box diagrams of Figure 9a.
For the SLAC measurement, one has $E=46$ GeV. Performing the calculation in the center of mass 
frame we obtain:
\bea
\frac{d\sigma^{\uparrow}}{d\Omega} -\frac{d\sigma^{\downarrow}}{d\Omega} &=& \frac{\alpha^3}{8}\frac{m}{t^2u^2\sqrt{s}}\sin\theta
\sqrt{1-\frac{4m^2}{s}} \nonumber \\
&\bigg[&3(s-4m^2)\bigg{(}t(u-s+2m^2)\ln{(\frac{-t}{s-4m^2})} \\
&& -u(t-s+2m^2)\ln{(\frac{-u}{s-4m^2})}\bigg{)}
-2(t-u)tu\bigg] \nonumber \\
\frac{d\sigma^{\uparrow}}{d\Omega} +\frac{d\sigma^{\downarrow}}{d\Omega}
&=& \frac{\alpha^2}{2st^2u^2}\bigg[(t^2+tu+u^2)^2+4m^2(m^2-t-u)
(t^2-tu+u^2)\bigg] \nonumber \ \ \ ,
\eea
\begin{figure}
    \begin{center}
    \includegraphics[angle=90,width=9cm]{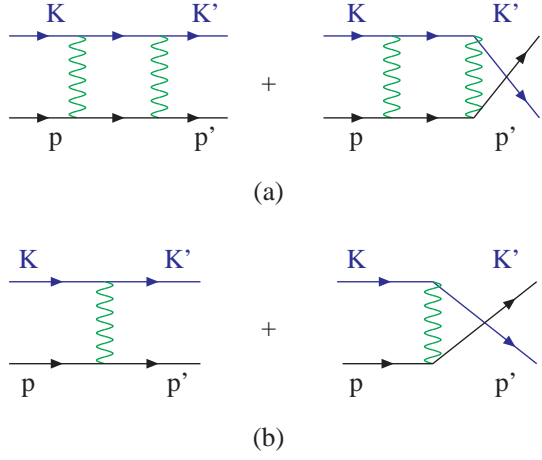}
    \end{center}
    \caption{(Color online) Diagrams contributing to the VAP for M\o ller scattering.}   
\end{figure}

Our results are in agreement with those of Refs. \cite{Barut:1960,DeRaad:1974,Dixon2004}\footnote{In Ref. \cite{Dixon2004}, ${\cal O}(\alpha^2)$ contributions arising from initial and  final state radiation effects were also computed. The corresponding contributions for the $ep$ VAP are smaller than the hadronic uncertainties arising at ${\cal O}(\alpha)$, so we do not consider them}. The resulting asymmetry is ploted in Fig. 10, and agrees with the corresponding figure in Ref. \cite{Dixon2004} (note that in Ref. \cite{Dixon2004}, the VAP is plotted vs. $\cos\theta$ rather than vs. $\theta$ as we do here).\newline

\begin{figure}
    \epsfxsize=4.0in
    \begin{center}
    \epsffile{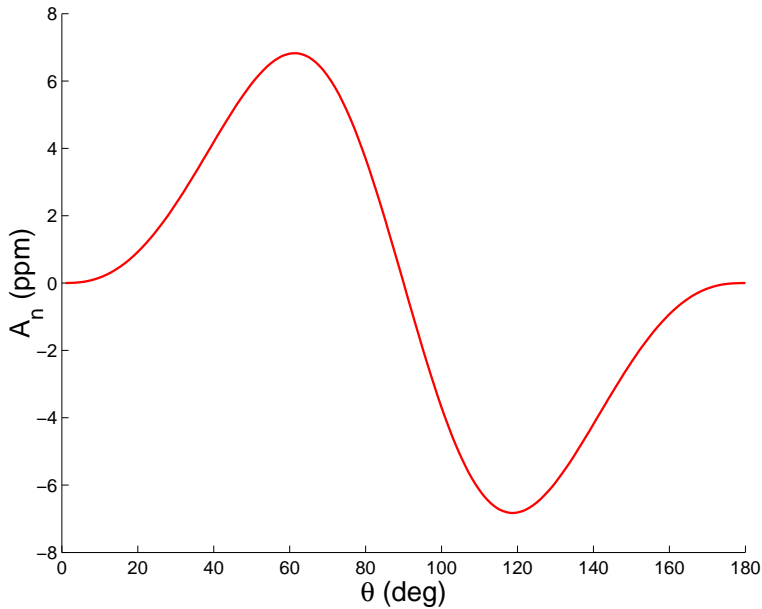}
    \end{center}
    \caption{(Color online) The M\o ller VAP vs CM scattering angle at the E158 kinematics.}   \end{figure}

\indent

\section{Conclusions}
\label{sec:conclusions}
\indent
\indent
In this study, we have computed the low-energy, backward angle VAP using an effective theory involving electrons, photons, and protons and have obtained a parameter-free prediction through ${\cal O}(p/M)^2$.
The VAP to this order is determined entirely by the imaginary part of the interference between the two-photon exchange, one-loop amplitude and the tree-level one-photon-exchange amplitude. In the limit that $M\to\infty$, our result exactly reproduces the VAP obtained in Ref. \cite{mott} for scattering from a structureless, infinitely heavy proton that over predicts the magnitude of $A_n$ at the kinematics of the SAMPLE experiment. We find that inclusion of all contributions through ${\cal O}(p/M)^2$ leads to agreement with experiment and leaves little room for important effects arising from dynamical pions or nucleon resonances at these energies. The leading counterterm contributions arise at ${\cal O}(p/M)^4$ and are consistent with zero. Thus, the SAMPLE measurement provides no evidence for unusual hadronic physics effects at these scales. The data also constrain the magnitude of the counterterm coefficients to be of natural size, and lead one to expect the VAP as measured by the A4 collaboration at Mainz to be at most of the order of a few ppm. Given the range of validity of our effective theory, however, we cannot produce a reliable prediction for VAP at the Mainz energies.

In this context, the results of the SAMPLE measurement have notable consequences for studies of weak interaction processes. In the case of both neutron $\beta$-decay and parity-violating $ep$ scattering, theoretical consideration of final state QED corrections to the leading-order weak amplitudes is important for the interpretation of various measurements \cite{McKeown:2002by}. To the extent that these measurements involve relatively low lepton energies, an analogous effective field theory computation of one-loop graphs involving the exchange of one weak vector boson and one photon should be reliable at the $\sim 20\% $ level relative to the size of other ${\cal O}(\alpha)$ corrections. Future, more precise measurements of the VAP at low-energies and over a range of angles would provide important tests of this provisional assessment. 

One might also ask how competitive the SAMPLE measurement is with other direct searches for new T-odd, P-even interactions. As discussed in Refs.\cite{Kurylov:2000ub,Ramsey-Musolf:1999nk}, direct searches are most relevant in symmetry-breaking scenarios wherein parity is broken at or above the scale for the breakdown of T.  Existing direct searches imply that $\alpha_T \lsim \ {\rm few}\ \times 10^{-3}$, where $\alpha_T$ is the ratio of a typical T-odd, P-even nuclear matrix element to those of the residual strong interaction. When translated into bounds on generic, dimension seven operator coefficients $C_{7}$ [under the normalization of Eq.  (\ref{eq:dimsevenrel})], one obtains $|C_{7}| \lsim 2$. The sensitivity of the SAMPLE  measurement is comparable. Given that conventional, hadronic final state effects that have been integrated out in our computation naturally imply a value of $C_{7a}$ with a magnitude of order unity, it appears unlikely that one will be able to circumvent the corresponding theoretical hadronic uncertainties as needed to make the VAP a direct probe of new physics. On the other hand, low-energy studies of $A_n$ could provide important information for the theoretical interpretation of other precision, electroweak observables. 

\vskip 0.5in

\noindent{\bf Acknowledgements}

It is a pleasure to thank C. Bauer, V. Cirigliano, B. Holstein, T. Ito, A. Kurylov, Y. Kolomensky, K. Kumar, F. Maas, D. Mack, C. Maekawa, R. D. McKeown, G. Prezeau, P.A. Souder, U. van Kolck, P. Vogel, and S.P. Wells for useful discussions. This work was supported in part under DOE contract DE-FG03-88ER40397 and NSF grant PHY-0071856.

\section*{Appendix A: Bremsstrahlung Computation}
\indent
\indent
Here, we show that the Bremsstrahlung amplitudes corresponding to Fig. 2 give a vanishing contribution to the VAP. The amplitudes are:
\begin{eqnarray}
{\cal M}^a &=&  \frac{-i}{q^2}\bar{u}(K^{\prime}) (ie) \gamma_{\mu} \frac{i(\nslash{K}-\nslash{l})+m)}{(K-l)^2-m^2}
(ie) \gamma^{\alpha} \epsilon_\alpha \frac{1+\gamma_5\nslash{S}}{2} u(K) \bar{u}(p^\prime) (ie) 
\gamma^{\mu} u(p) \nonumber \\
{\cal M}^{b} &=& \frac{-i}{q^2} \bar{u}(K^{\prime}) (ie) \gamma^{\alpha} \epsilon_\alpha 
\frac{i(\nslash{K}^{\prime}+\nslash{l})+m)}{(K^\prime+l)^2-m^2}
(ie) \gamma_{\mu} \frac{1+\gamma_5\nslash{S}}{2} u(K) \bar{u}(p^\prime) (ie) 
\gamma_{\mu} u(p) \nonumber \\
{\cal M}^{c} &=& \frac{-i}{q^2} \bar{u}(K^{\prime}) (ie) \gamma^{\mu}  u(K) \bar{u}(p^\prime) (ie) \gamma^{\mu} \frac{i(\nslash{p}^{\prime}+\nslash{l})+M)}{(p^{\prime}+l)^2-M^2}
(ie) \gamma^{\alpha} \epsilon_\alpha u(p) \nonumber \\
{\cal M}^{d} &=& \frac{-i}{q^2}  \bar{u}(K^\prime) (ie) \gamma^{\mu}  u(K) \bar{u}(p^\prime) (ie) \gamma^{\alpha}\epsilon_\alpha 
\frac{i(\nslash{p}-\nslash{l})+M)}{(p-l)^2-M^2}
(ie) \gamma^{\mu} u(p) 
\end{eqnarray}
Here, $l_\mu$ is the radiated photon momentum. The square of the invariant amplitude
\be
{\cal M^B} = \Bigl\vert {\cal M}^a+\cdots +{\cal M}^d \Bigr\vert^2
\ee
depends on ten different products of leptonic and hadronic tensors. The leptonic tensors are:
\begin{eqnarray}
L^{aa}_{\mu\nu} &=& Tr[(\nslash{K}^{\prime}+m)\gamma_{\mu}\frac{(\nslash{K}-\nslash{l}+m)}{(K-l)^2-m^2}\gamma_\alpha
\frac{1+\gamma_5\nslash{S}}{2}(\nslash{K}+m)\gamma_{\beta}\frac{(\nslash{K}-\nslash{l}+m)}{(K-l)^2-m^2}\gamma_{\nu}]
\epsilon^\alpha\epsilon^{* \beta} \nonumber \\
L^{ab}_{\mu\nu} &=& Tr[(\nslash{K}^{\prime}+m)\gamma_{\mu}\frac{(\nslash{K}-\nslash{l}+m)}{(K-l)^2-m^2}\gamma_\alpha
\frac{1+\gamma_5\nslash{S}}{2}(\nslash{K}+m)\gamma_{\nu}\frac{( \nslash{K}^{\prime}+\nslash{l}+m)}{(K^\prime+l)^2-m^2}\gamma_{\beta}]
\epsilon^\alpha\epsilon^{* \beta} \nonumber \\
L^{ac}_{\mu\nu} &=& Tr[(\nslash{K}^{\prime}+m)\gamma_{\mu}\frac{(\nslash{K}-\nslash{l}+m)}{(K-l)^2-m^2}\gamma_\alpha
\frac{1+\gamma_5\nslash{S}}{2}(\nslash{K}+m)\gamma_{\nu}]
\epsilon^\alpha \nonumber \\
L^{ad}_{\mu\nu} &=& L^{ac}_{\mu\nu} \nonumber \\
L^{bb}_{\mu\nu} &=& Tr[(\nslash{K}^{\prime}+m)\gamma_\alpha 
\frac{(\nslash{K}^{\prime}+\nslash{l}+m)}{(K^\prime+l)^2-m^2}\gamma_{\mu}
\frac{1+\gamma_5\nslash{S}}{2}(\nslash{K}+m)\gamma_{\nu} \frac{(\nslash{K}^{\prime}+\nslash{l}+m)}{(K^\prime+l)^2-m^2}\gamma_{\beta}]
\epsilon^\alpha\epsilon^{* \beta} \nonumber \\
L^{bc}_{\mu\nu} &=& Tr[(\nslash{K}^{\prime}+m)\gamma_\alpha\frac{(\nslash{K}^{\prime}+\nslash{l}+m)}{(K^\prime+l)^2-m^2}\gamma_{\mu}
\frac{1+\gamma_5\nslash{S}}{2}(\nslash{K}+m)\gamma_{\nu}] 
\epsilon^\alpha \nonumber \\
L^{bd}_{\mu\nu} &=& L^{bc}_{\mu\nu} \nonumber \\
L^{cc}_{\mu\nu} &=& Tr[(\nslash{K}^{\prime}+m)\gamma_{\mu}
\frac{1+\gamma_5\nslash{S}}{2}(\nslash{K}+m)\gamma_{\nu}] \nonumber \\
L^{cd}_{\mu\nu} &=& L^{cc}_{\mu\nu} \nonumber \\
L^{dd}_{\mu\nu} &=& L^{cc}_{\mu\nu} 
\end{eqnarray}
The corresponding hadronic tensors are:
\begin{eqnarray}
H_{aa}^{\mu\nu} &=& Tr[(\nslash{p}^{\prime}+M)\gamma^{\mu}
(\nslash{p}+M)\gamma^{\nu}] \nonumber \\
H_{ab}^{\mu\nu} &=& H_{aa}^{\mu\nu} \nonumber \\
H_{ac}^{\mu\nu} &=& Tr[(\nslash{p}^{\prime}+M)\gamma^{\mu}
(\nslash{p}+M)\gamma^{\beta}\frac{(\nslash{p}-\nslash{l}+M)}{(p-l)^2-M^2}\gamma^{\nu}]
\epsilon^*_\beta \nonumber \\
H_{ad}^{\mu\nu} &=& Tr[(\nslash{p}^{\prime}+M)\gamma^{\mu}
(\nslash{p}+M)\gamma^{\nu}\frac{(\nslash{p^\prime}+\nslash{l}+M)}{(p^\prime+l)^2-M^2}\gamma^{\beta}]
\epsilon^*_\beta \nonumber \\
H_{bb}^{\mu\nu} &=& H_{aa}^{\mu\nu} \nonumber \\
H_{bc}^{\mu\nu} &=& H_{aa}^{\mu\nu} \nonumber \\
H_{bd}^{\mu\nu} &=& H_{ad}^{\mu\nu}  \nonumber \\
H_{cc}^{\mu\nu} &=& Tr[(\nslash{p}^{\prime}+M)\gamma^{\alpha}\frac{(\nslash{p}-\nslash{l}+M)}{(p-l)^2-M^2}\gamma^{\mu}
(\nslash{p}+M)\gamma^{\nu}\frac{(\nslash{p}-\nslash{l}+M)}{(p-l)^2-M^2}\gamma^{\beta}]
\epsilon_\alpha\epsilon^*_\beta \nonumber \\
H_{cd}^{\mu\nu} &=& Tr[(\nslash{p}^{\prime}+M)\gamma^{\alpha}\frac{(\nslash{p}-\nslash{l}+M)}{(p-l)^2-M^2}\gamma^{\mu}
(\nslash{p}+M)\gamma^{\beta}\frac{(\nslash{p^\prime}+\nslash{l}+M)}{(p^\prime+l)^2-M^2}\gamma^{\nu}] 
\epsilon_\alpha\epsilon^*_\beta\nonumber \\
H_{dd}^{\mu\nu} &=& Tr[(\nslash{p}^{\prime}+M)\gamma^{\mu}\frac{(\nslash{p^\prime}+\nslash{l}+M)}{(p^\prime+l)^2-M^2}\gamma^{\alpha}
(\nslash{p}+M)\gamma^{\beta}\frac{(\nslash{p^\prime}+\nslash{l}+M)}{(p^\prime+l)^2-M^2}\gamma^{\nu}] \epsilon_\alpha\epsilon^*_\beta
\end{eqnarray}
We now need to compute:
\begin{eqnarray}
\label{brem-int}
\cal M^B &=& \sum_{\rm pol}\int{d^4l}\Biggl\{\frac{1}{q^4}\Biggl[L^{aa}_{\mu\nu} H_{aa}^{\mu\nu}+
L^{ab}_{\mu\nu} H_{ab}^{\mu\nu}
+L^{ac} H_{ac}^{\mu\nu}  
+L^{ad}_{\mu\nu} H_{ad}^{\mu\nu} 
+L^{bb}_{\mu\nu} H_{bb}^{\mu\nu} \nonumber \\
&+&L^{bc}_{\mu\nu} H_{bc}^{\mu\nu} 
+L^{bd}_{\mu\nu} H_{bd}^{\mu\nu} 
+ L^{cc}_{\mu\nu} H_{cc}^{\mu\nu} 
+L^{cd}_{\mu\nu} H_{cd}^{\mu\nu}
+L^{dd}_{\mu\nu} H_{dd}^{\mu\nu}\Biggr]+{\rm h.c.}\Biggr\} \nonumber \\
&=& \sum_{\rm pol}\int{d^4l}\Biggl\{\frac{1}{q^4}\Biggl[ (H_{ac}^{\mu\nu}+H_{ad}^{\mu\nu})
(L^{ac}_{\mu\nu}+L^{ad}_{\mu\nu})+
H_{aa}^{\mu\nu}(L^{aa}_{\mu\nu}+L^{ab}_{\mu\nu}+L^{bb}_{\mu\nu}) \nonumber \\
&+&L_{cc}^{\mu\nu}(H^{cc}_{\mu\nu}+H^{cd}_{\mu\nu}+H^{dd}_{\mu\nu})\Biggr]+{\rm h.c.}\Biggr\}\ \ \ ,
\end{eqnarray}
where the sum is over all polarizations of the radiated photon.
We are only interested in the terms proportional to \( \epsilon_{\alpha\beta\gamma\delta}S^\alpha k^\beta
k^{\prime\gamma}p^\delta\). First we investigate the momentum integrals:
\begin{eqnarray}
\label{twopoint}
i\pi^2\cal I^B = \int&{d^4l}&\bigg[\frac{1}{(p^\prime+l)^2-M^2}\frac{1}{(p^\prime+l)^2-M^2}
+\frac{1}{(p^\prime+l)^2-M^2}\frac{1}{(p-l)^2-M^2} \nonumber \\
&+&\frac{1}{(p-l)^2-M^2}\frac{1}{(p-l)^2-M^2}+
\frac{1}{(k^\prime+l)^2-m^2}\frac{1}{(k^\prime+l)^2-m^2}  \nonumber \\
&+& \ldots \bigg]
\end{eqnarray}
We can evaluate the generic two point integral as defined by:
\begin{eqnarray}
i\pi^2 B(p^2;m_1^2,m_2^2) = \mu^{4-n}\int&{d^nq}&\bigg[\frac{1}{q^2+m_1^2-i\epsilon}\times
\frac{1}{(q+p)^2+m_2^2-i\epsilon}\bigg]
\end{eqnarray}
We are only interested in the imaginary part of $B$. We find that 
above the physical treshold $s=-p^2\ge(m_1+m_2)^2$ this integral develops an imaginary part Ref. \cite{Bardin}:
\begin{eqnarray}
{\rm Im} B(p^2;m_1^2,m_2^2) = \pi\frac{\sqrt{\lambda(s,m_1^2,m_2^2)}}{s}\Theta(s-(m_1+m_2)^2)
\end{eqnarray}
Evaluating the B functions for the kinematics involved here we find that none of the 
integrals of Eqn. (\ref{twopoint}) develop an imaginary part. As such
evaluating the traces and performing the integration we obtain a rezult of the form:
\begin{eqnarray}
\cal M^B &=& f_1(m,M,s,t,u) +f_2(m,M,s,t,u)i\epsilon_{\alpha\beta\gamma\delta}S^\alpha k^\beta
k^{\prime\gamma}p^\delta + h.c. \nonumber \\
 &=&2f_1(m,M,s,t,u)
\end{eqnarray}
Hence, we find no contribution to $A_n$.

\section*{Appendix B: Local Operators}
As discussed in the text, we are interested in computing the contribution to the VAP from local, four fermion $eeNN$ operators.  The lowest dimension operators of this form have dimension six. First, we show by explicit calculation that all $d=6$ operators give vanishing contributions to $A_n$.
The most general form for the $d=6$ operators are 
\begin{eqnarray}
{\cal O}_{eN}^{6a} & = & 
\frac{\alpha^2}{M^2}\bar{e}(C_1+C_2\gamma_5)e\bar{N}(C_1^\prime+C_2^\prime\gamma_5)N \nonumber \\
{\cal O}_{eN}^{6b} & = & 
\frac{\alpha^2}{M^2}\bar{e}(C_3+C_4\gamma_5)\gamma^\mu e
\bar{N}(C_3^\prime+C_4^\prime\gamma_5)\gamma^\mu\\
{\cal O}_{eN}^{6c} & = & 
\frac{\alpha^2}{M^2}\bar{e}(C_5+C_6\gamma_5)\sigma^{\mu\nu} e\bar{N}(C_5^\prime+C_6^\prime\gamma_5)\sigma_{\mu\nu} N 
\end{eqnarray}
where we have used relativistic nucleon fields $N$ (the corresponding argument carries over straightforwardly in the heavy baryon formalism). 
To make the above hermitian we require all the constants $C_{eN}^i$ to be real. We now compute the interference of the amplitudes associated with these operators and 
the tree amplitude ${\cal M}_\gamma$, retaining only the desired structure $\epsilon_{\alpha\beta\gamma\delta}
S^\alpha p^\beta K^\gamma K^{\prime\delta}$. 
The corresponding leptonic and hadronic tensors are
\begin{eqnarray}
L^\mu_{6a} &=& Tr[(\nslash{K}^\prime+m)(C_1+C_2\gamma_5)\frac{1+\gamma_5\nslash{S}}{2}(\nslash{K}+m)\gamma^\mu] \nonumber \\
L^{\mu\nu}_{6b} &=& Tr[(\nslash{K}^\prime+m)(C_3+C_4\gamma_5)\gamma^\nu
\frac{1+\gamma_5\nslash{S}}{2}(\nslash{K}+m)\gamma^\mu] \nonumber \\
L^{\mu\nu\alpha}_{6c} &=& Tr[(\nslash{K}^\prime+m)(C_5+C_6\gamma_5)\sigma^{\nu\alpha}
\frac{1+\gamma_5\nslash{S}}{2}(\nslash{K}+m)\gamma^\mu] \nonumber \\
H^\mu_{6a} &=& Tr[(\nslash{p}^\prime+m)(C_1^\prime+C_2^\prime\gamma_5)(\nslash{p}+m)\gamma^\mu] \nonumber \\
H^{\mu\nu}_{6b} &=& Tr[(\nslash{p}^\prime+m)(C_3^\prime+C_4^\prime\gamma_5)\gamma^\mu(\nslash{p}+m)\gamma^\mu] \nonumber \\
H^{\mu\nu\alpha}_{6c} &=& Tr[(\nslash{p}^\prime+m)(C_5^\prime+C_6^\prime\gamma_5)\sigma^{\mu\alpha}(\nslash{p}+m)\gamma^\mu] \nonumber \\
{\cal M}_{6}{\cal M}_\gamma^{\ast} + {\rm h.c.} &=& \frac{(4\pi\alpha)\alpha^2}{tM^2}
\left[L^\mu_{6a} H_{\mu(6a)}+L^{\mu\nu}_{6b}H_{\mu\nu(6b)}+
L^{\mu\nu\alpha}_{6c}H_{\mu\nu\alpha(6c)}\right]+{\rm, h.c.}
\end{eqnarray}
Evaluating the traces and keeping only the terms of interest we obtain
\begin{eqnarray}
{\cal M}_{6}{\cal M}_\gamma^{\ast} + {\rm h.c.} &=& i16\frac{(4\pi\alpha)\alpha^2}{tM^2}
(C_1C_1^\prime M - C_4C_4^\prime m)
\epsilon_{\alpha\beta\gamma\delta}S^\alpha p^\beta K^\gamma K^{\prime\delta}+{\rm h.c.} 
\end{eqnarray}
Since all the C's are real we see there is no contribution from dimmension six terms. This results is as expected, as the operators ${\cal O}_{6a-c}$ are even under both T and P.

Now consider $d=7$ operators. As for the $d=6$ operators, all contributions
from T-even P-even $d=7$ operators will vanish. We may, however, write down two
Hermitian T-odd, P-even $d=7$ operators:
\begin{eqnarray}
{\cal O}_{eN}^{7a} & = & \frac{\alpha^2}{M^3}C_{7a}\bar{e}\gamma_5\sigma^{\mu \nu}({\overleftarrow D}+{\overrightarrow D})_\nu \bar{N}\gamma_5\gamma\mu N\\
{\cal O}_{eN}^{7b} & = & \frac{\alpha^2}{M^3}C_{7b}\bar{e}\gamma_5\gamma_\mu e\bar{N}\gamma_5\sigma^{\mu\nu}({\overleftarrow D}+{\overrightarrow D})_\nu N 
\end{eqnarray}
As before we evaluate the interference of the above with ${\cal M}_\gamma$. The corresponding leptonic and hadronic tensors are:
\begin{eqnarray}
L_{7a}^{\mu\nu} &=& iTr[(\nslash{K}^\prime+m)\gamma_5\sigma^{\mu \alpha}q_\alpha\frac{1+\gamma_5\nslash{S}}{2}(\nslash{K}+m)\gamma^\nu] \nonumber \\
L_{7b}^{\mu\nu} &=& Tr[(\nslash{K}^\prime+m)\gamma_5\gamma^\mu\frac{1+\gamma_5\nslash{S}}{2}(\nslash{K}+m)
\gamma^\nu]
\nonumber \\
H^{\mu\nu}_{7a} &=& Tr[(\nslash{p}^\prime+m)\gamma_5\gamma^\mu (\nslash{p}+m)\gamma^\nu] \nonumber \\
H^{\mu\nu}_{7b} &=& iTr[(\nslash{p}^\prime+m)\gamma_5\sigma^{\mu\alpha}q^a(\nslash{p}+m)\gamma^\nu] \nonumber \\
{\cal M}_{7}{\cal M}_\gamma^{\ast} +{\rm  h.c.} &=& i\frac{(4\pi\alpha)\alpha^2}{tM^3}\left[C_{7a}L_{7a}^{\mu\nu}H_{\mu\nu (7a)}
+C_{7b}L_{7b}^{\mu\nu}H_{\mu\nu (7b)}\right]+ {\rm h.c.}
\end{eqnarray}
Evaluating the traces we note that only the $L_{7a}^{\mu\nu}H_{\mu\nu (7a)}$ contributes:
\begin{eqnarray}
{\cal M}_{7}{\cal M}_\gamma + {\rm h.c.} &=& \frac{16(4\pi\alpha)\alpha^2C_{7a}}{M^3}
\epsilon_{\alpha\beta\gamma\delta}
S^\alpha p^\beta k^\gamma k^{\prime\delta}
\end{eqnarray}
We are intrested in the contribution such a term gives to the VAP. Keep only the leading piece of the
tree amplitude we get:
\begin{eqnarray}
A_n^{(7)} = \frac{\alpha C_{7a}}{4\pi}{t^2 |\vec{K}| |\vec{K^\prime}| \sin\theta\over M^2[8M^2E^2+2(2E+M)tM+t^2]}
\end{eqnarray}

\section*{Appendix C: Loop Integrals}
Here, we provide additional details about the computation of ${\cal M}_{\gamma\gamma}$.
As noted in the text, the contribution from the crossed-box diagram vanishes, so we consider only 
${\rm Im} {\cal M}_{\gamma\gamma}^{\rm box}{\cal M}_{\gamma}^{\ast}$. We may express the latter in terms of the leptonic and hadronic tensors:

\begin{figure}
    \begin{center}
    \includegraphics[angle=90,width=9cm]{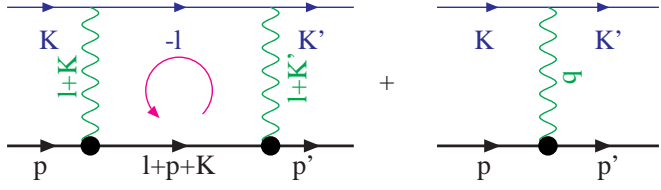}
    \end{center}
    \caption{(Color online) Momentum routing for the $\gamma\gamma$ box graph integrals.}   
\end{figure}

\begin{eqnarray}
L^{\mu\nu\alpha} &=&  \bar{u}(K^{\prime}) (ie) \gamma^{\mu} \frac{i(-\nslash{l}+m)}{l^2-m^2}
(ie) \gamma^{\nu} \frac{1+\gamma_5 \nslash{S}}{2} u(K) \bar{u}(K) (ie) \gamma^\alpha u(K^{\prime}) \nonumber \\
H_{\mu\nu\alpha} &=&  \bar{u}(p^{\prime})[ie(1+r(l+K^{\prime})^2)\gamma_{\mu}-\frac{\kappa\sigma_{\mu\beta}}{2M}(l+K^{\prime})^\beta] 
\frac{i(\nslash{l}+\nslash{K^{\prime}}+\nslash{p^{\prime}}+M)}
{((l+K^{\prime}+p^{\prime})^2-M^2} \nonumber \\ 
&\times&[ie(1+r(l+K)^2)\gamma_{\nu}+\frac{\kappa\sigma_{\nu\delta}}{2M}(l+K)^\delta] u(p)\nonumber \\
&\times&\bar{u}(p)[ie(1+r(K-K^{\prime})^2)\gamma_\alpha +\frac{\kappa\sigma_{\alpha\gamma}}{2M}(K-K^{\prime})^\gamma]u(p^{\prime}) \nonumber \\
 {\cal M}_{\gamma\gamma}^{\rm box}{\cal M}_{\gamma}^{\ast}&=& \int \frac{d^4l}{(2\pi)^2} L^{\mu\nu\alpha} \frac{-i}{(l+K^{\prime})^2} \frac{-i}{(l+K)^2}
\frac{-i}{(K-K^{\prime})^2} H_{\mu\nu\alpha} 
\end{eqnarray}
where
\bea
r=R-1
\eea
We define the loop integrals from above as follows:
\begin{eqnarray}
i\pi^2D_0 &=& \int d^4l \frac{1}{(l^2-m^2)(l+K^{\prime})^2[(l+K^{\prime}+p^{\prime})^2-M^2](l+K)^2} \nonumber \\
i\pi^2D_{\alpha} &=& \int d^4l \frac{l_{\alpha}}{(l^2-m^2)(l+K^{\prime})^2[(l+K^{\prime}+p^{\prime})^2-M^2](l+K)^2} \nonumber \\
i\pi^2D_{\alpha\beta} &=& \int d^4l \frac{l_{\alpha}l_{\beta}}{(l^2-m^2)(l+K^{\prime})^2
[(l+K^{\prime}+p^{\prime})^2-M^2](l+K)^2} \nonumber \\
i\pi^2D_{\alpha\beta\gamma} &=& \int d^4l \frac{l_{\alpha}l_{\beta}l_{\gamma}}{(l^2-m^2)(l+K^{\prime})^2
[(l+K^{\prime}+p^{\prime})^2-M^2](l+K)^2} \nonumber \\
i\pi^2D_{\alpha\beta\gamma\delta} &=& \int d^4l \frac{l_{\alpha}l_{\beta}l_{\gamma}l_{\delta}}{(l^2-m^2)(l+K^{\prime})^2
[(l+K^{\prime}+p^{\prime})^2-M^2](l+K)^2}
\end{eqnarray}
In order to evaluate these integrals, we follow the methods of Refs. \cite{thft,pave}, and our notation follows that of Ref. \cite{pave}. To this end, we need to compute
the following three point functions
\begin{eqnarray}
i\pi^2C_0(1,2,3) &=& \int d^4l \frac{1}{(l^2-m^2)(l+K^{\prime})^2[(l+K^{\prime}+p^{\prime})^2-M^2]} \nonumber \\
i\pi^2C_0(1,2,4) &=& \int d^4l \frac{1}{(l^2-m^2)(l+K^{\prime})^2(l+K)^2} \nonumber \\
i\pi^2C_0(1,3,4) &=& \int d^4l \frac{1}{(l^2-m^2)[(l+K^{\prime}+p^{\prime})^2-M^2](l+K)^2} \nonumber \\
i\pi^2C_0(2,3,4) &=& \int d^4l \frac{1}{(l+K^{\prime})^2[(l+K^{\prime}+p^{\prime})^2-M^2](l+K)^2} \nonumber \\
\end{eqnarray}
and two point functions
\begin{eqnarray}
i\pi^2B_0(1,2) &=& \int d^4l \frac{1}{(l^2-m^2)(l+K^{\prime})^2} \nonumber \\
i\pi^2B_0(1,3) &=& \int d^4l \frac{1}{(l^2-m^2)[(l+K^{\prime}+p^{\prime})^2-M^2]} \nonumber \\
i\pi^2B_0(1,4) &=& \int d^4l \frac{1}{(l^2-m^2)(l+K)^2} \nonumber \\
i\pi^2B_0(2,4) &=& \int d^4l \frac{1}{(l+K^{\prime})^2(l+K)^2} \nonumber \\
i\pi^2B_0(2,3) &=& \int d^4l \frac{1}{(l+K^{\prime})^2[(l+K^{\prime}+p^{\prime})^2-M^2]} \nonumber \\
i\pi^2B_0(3,4) &=& \int d^4l \frac{1}{[(l+K^{\prime}+p^{\prime})^2-M^2](l+K)^2} \nonumber \\
\end{eqnarray}
For all the B,C and D integrals above we are interested only in the imaginary part. The only two-, three- and 
four-point integrals with non-vanishing imaginary parts are:

\begin{eqnarray}
{\rm Im}\ D_0 &=& \frac{2\pi}{-t} \ln(\frac{-t}{\lambda^2})\frac{1}{\sqrt{\Lambda}}
\Theta(s-(m+M)^2)  \nonumber \\
{\rm Im}\ C_0(1,2,3) &=& \frac{\pi}{\sqrt{\Lambda}} \ln(\frac{\Lambda}{s\lambda^2})
\Theta(s-(m+M)^2)  \nonumber \\
{\rm Im}\ C_0(1,3,4) &=& {\rm Im}[C_0(1,2,3)] = C_0 \nonumber \\
{\rm Im}\ B_0(1,3) &=& \pi\frac{\sqrt{\Lambda}}{s}\Theta(s-(m+M)^2)
\end{eqnarray}
In the above \( \lambda \) is the photon mass and
\( \Lambda = s^2-2s(M^2+m^2)+(M^2-m^2)^2 \). 

Although space considerations preclude a complete delineation of the calculation here, it is instructive to consider in more detail the evaluation of one of the four-point integrals required. Specifically, we consider
\begin{eqnarray}
D^\alpha = p_1^\alpha D_{11} + p_2^\alpha D_{12} + p_3^\alpha D_{13}
\end{eqnarray}
For the kinematics considered here the Passarino and Veltman momenta and masses are:
\begin{eqnarray}
\begin{array}{cc}
p_1 = K & m_1 = m \\
p_2 = p & m_2 = 0 \\
p_3 = -p^\prime & m_3 = M \\
p_4 = -K^\prime & m_4 = 0 
\end{array}
\end{eqnarray}
We then have for the ${\rm Im}\ D_{ij}$
\begin{eqnarray}
{\rm Im}\left( \begin{array}{c}
D_{11} \\ D_{12} \\ D_{13}
\end{array} \right)  = X^{-1}{\rm Im}
\left( \begin{array}{c}
R_{20} \\ R_{21} \\ R_{22}
\end{array} \right) 
\end{eqnarray}
where 
\begin{eqnarray}
R_{20} &=& \frac{1}{2}[f_1D_0+C_0(1,3,4)-C_0(2,3,4)]=\frac{1}{2}(2D_0m^2+C_0) \nonumber \\
R_{21} &=& \frac{1}{2}[f_2D_0+C_0(1,2,4)-C_0(1,3,4)]=\frac{1}{2}[2D_0(s-M^2-m^2)-C_0] \nonumber \\
R_{22} &=& \frac{1}{2}[f_3D_0+C_0(1,2,3)-C_0(1,2,4)]=\frac{1}{2}[-2D_0(s-M^2-m^2)+C_0] \ \ \ ,
\end{eqnarray}
where
\begin{eqnarray}
f_1 &=& m_1^2-m_2^2-p^2_1 = 2m^2 \nonumber \\
f_2 &=& m_1^2-m_2^2+p_1^2-p_5^2 = (s-M^2-m^2) \nonumber \\
f_3 &=& m_2^2 - m_4^2-p_4^2+p_5^2 = -f_2 \ \ \ ,
\end{eqnarray}
and where the inverse of the momentum matrix X is:
\begin{eqnarray}
X^{-1} &=& 
\left( \begin{array}{ccc}
p_1^2 & p_1p_2 & p_1p_3 \\
p_1p_2 & p_2^2 & p_2p_3 \\
p_1p_3 & p_2p_3 & p_3^2 
\end{array} \right)^{-1} \nonumber \\
&=&\left( \begin{array}{ccc}
\frac{4M^2-t}{\Lambda+ts} & \frac{3M^2+m^2-s-t}{\Lambda+ts} & \frac{M^2-m^2+s}{\Lambda+ts} \\
\frac{3M^2+m^2-s-t}{\Lambda+ts} & \frac{2(M^2+s+t)m^2-(s+t-M^2)^2-m^4}{t(\Lambda+ts)} & 
\frac{M^2-m^2}{\Lambda+ts}-\frac{1}{t} \\
\frac{M^2-m^2+s}{\Lambda+ts} & \frac{M^2-m^2}{\Lambda+ts}-\frac{1}{t} & 
\frac{s}{\Lambda+ts}-\frac{1}{t}
\end{array} \right) 
\end{eqnarray}
After performing the necessary algebra we obtain
\begin{eqnarray}
{\rm Im}[D_{11}] &=& -\frac{D_0[2((m-M)^2-s)((m+M)^2-s)+(m^2-M^2+s)t]-2C_0(s+M^2-m^2)}{2(\Lambda+ts)} \nonumber \\
{\rm Im}[D_{12}] &=& -\frac{D_0(m^4+(t-2(M^2+s))+(M^2-s)(M^2-s-t))m^2+2C_0(m^2-M^2)}{2(\Lambda+ts)} \nonumber \\
{\rm Im}[D_{13}] &=& \frac{-D_0\Lambda+2C_0s}{2(\Lambda+ts)}\ \ \ .
\end{eqnarray}
Similar steps are required in evaluating the other four-point integrals.


\begin{thebibliography} {999}

\bibitem{Lising:2000pa}
L.~J.~Lising {\it et al.}  [emiT Collaboration],
Phys.\ Rev.\ C {\bf 62}, 055501 (2000)
[arXiv:nucl-ex/0006001].

\bibitem{Herczeg1995} P. Herczeg, \lq\lq Time Reversal Violation in Nuclear Processes", in {\em Symmetries and Fundamental Interactions in Nuclei}, W.C. Haxton and E.M. Henley, eds, World Scientific (1995), p. 89 and references therein.

\bibitem{Boehm1995} F. Boehm, \lq\lq Time Reversal Tests in Nuclei", in {\em Symmetries and Fundamental Interactions in Nuclei}, W.C. Haxton and E.M. Henley, eds, World Scientific (1995), p.67 and references therein.
\bibitem{Kurylov:2000ub}
A.~Kurylov, G.~C.~McLaughlin and M.~J.~Ramsey-Musolf,
Phys.\ Rev.\ D {\bf 63}, 076007 (2001)
[arXiv:hep-ph/0011185].

\bibitem{Ramsey-Musolf:1999nk}
M.~J.~Ramsey-Musolf,
Phys.\ Rev.\ Lett.\  {\bf 83}, 3997 (1999)
[Erratum-ibid.\  {\bf 84}, 5681 (2000)]
[arXiv:hep-ph/9905429].

\bibitem{Engel:1995vv}
J.~Engel, P.~H.~Frampton and R.~P.~Springer,
Phys.\ Rev.\ D {\bf 53}, 5112 (1996)
[arXiv:nucl-th/9505026].

\bibitem{Conti:xn}
R.~S.~Conti and I.~B.~Khriplovich,
Phys.\ Rev.\ Lett.\  {\bf 68} (1992) 3262.

\bibitem{Khriplovich:1990ef}
I.~B.~Khriplovich,
Nucl.\ Phys.\ B {\bf 352}, 385 (1991).

\bibitem{Davis1980} B.R. Davis, S.E. Koonin, and P. Vogel, Phys. Rev. C {\bf 22}, 1233 (1980).

\bibitem{trans-ex} S.P. Wells {\it et al.}, SAMPLE Collaboration, Phys. 
Rev. C {\bf 63}, 064001 (2001).

\bibitem{mott} N.F. Mott, Proc. Roy. Soc. (London) {\bf A 135}, 429 
(1932) .

\bibitem{Blunden:2003sp}
P.~G.~Blunden, W.~Melnitchouk and J.~A.~Tjon,
Phys.\ Rev.\ Lett.\  {\bf 91}, 142304 (2003)
[arXiv:nucl-th/0306076].

\bibitem{McKeown:2002by}
R.~D.~McKeown and M.~J.~Ramsey-Musolf,
Mod.\ Phys.\ Lett.\ A {\bf 18}, 75 (2003)
[arXiv:hep-ph/0203011].

\bibitem{guich} P. A. M. Guichon and M. Vanderhaeghen, Prog. Part. 
Nucl. Phys. {\bf 41} 125 (1998).


\bibitem{Maas2003} F. Maas, talk given at ECT$\ast$, Trento, Italy, April 2004, and F. Maas, private communication.

\bibitem{E158} P. L. Anthony et al. [SLAC E158 Collaboration], hep-ex/0312035; K. Kumar and
Y. Kolomensky, private communication.

\bibitem{Barut:1960} A.O. Barut and C. Fronsdal, Phys. Rev. {\bf 120}, 1871 (1960).

\bibitem{DeRaad:1974} L.L. DeRaad and Y. J. Ng, Phys. Rev. {\bf D10}, 683 (1974); Phys. Rev. {\bf D 10}, 3440 (1974); Phys. Rev. {\bf D 11}, 1586 (1975). 

\bibitem{Dixon2004} L. Dixon, M. Schreiber, hep-ph/040222.

\bibitem{Simon:hu}
G.~G.~Simon, C.~Schmitt, F.~Borkowski and V.~H.~Walther,
Nucl.\ Phys.\ A {\bf 333}, 381 (1980).

\bibitem{Bhaduri} R.K. Bhaduri, \lq\lq Models of the Nucleon", Addison-Wesley, Redwood City, CA (1988). 


\bibitem{DeRujula:1972te}
A.~De Rujula, J.~M.~Kaplan and E.~De Rafael,
Nucl.\ Phys.\ B {\bf 35}, 365 (1971).

\bibitem{thft} G. 't Hooft and M. Veltman, Nucl. Phys. {\bf B 153}, 365-401 (1979) .
\bibitem{pave} G. Passarino and M. Veltman, Nucl. Phys. {\bf B 160}, 151-207 (1979) .
\bibitem{Zhu:2000zf}
S.~L.~Zhu, S.~Puglia and M.~J.~Ramsey-Musolf,
Phys.\ Rev.\ D {\bf 63}, 034002 (2001)
[arXiv:hep-ph/0009159].
\bibitem{vann} P. van Nieuwenhuizen, Nucl. Phys. {\bf B 28}, 429-454 (1971) .


\bibitem{Afanasev:2002gr}
A.~Afanasev, I.~Akushevich and N.~P.~Merenkov,
arXiv:hep-ph/0208260.


\bibitem{Chen:2004tw}
Y.~C.~Chen, A.~Afanasev, S.~J.~Brodsky, C.~E.~Carlson and M.~Vanderhaeghen,
arXiv:hep-ph/0403058.


\bibitem{Pasquini:bw}
B.~Pasquini, D.~Drechsel, M.~Vanderhaeghen, M.~Gorchtein and A.~Metz,
AIP Conf.\ Proc.\  {\bf 675} (2003) 646.

\bibitem{Bernabeu} J. Bernabeu and J.A. Penarrocha, Phys. Rev. {\bf D22}, 1082 (1980).


\bibitem{Bardin} D. Bardin, Lectures at the European School of High Energy Physics, Slovakia Aug.-Sept. 1999

\bibitem{Pasquini:2004pv}
B.~Pasquini and M.~Vanderhaeghen,
arXiv:hep-ph/0405303.

\bibitem{Afanasev:2004hp}
A.~V.~Afanasev and N.~P.~Merenkov,
arXiv:hep-ph/0406127.

\bibitem{Afanasev:2004pu}
A.~Afanasev and N.~P.~Merenkov,
arXiv:hep-ph/0407167.



\end{thebibliography}
\end{document}